\documentclass[a4paper,fleqn,usenatbib]{mnras}

\usepackage{times}

\usepackage[T1]{fontenc}
\usepackage{ae,aecompl}

\usepackage{graphicx}	
\usepackage{amsmath}	
\usepackage{amssymb}	

\usepackage{graphicx}
\usepackage{amssymb}
\usepackage{lscape}
\usepackage{ulem}
\usepackage{txfonts}
\usepackage{url}
\usepackage{subfig}

\newcommand\mm{$\mu$m}
\newcommand\herschel{\textit{Herschel}}
\newcommand\hermescollab{HerMES Collaboration~}

\title[Multiplicity of bright \textit{Herschel} sources]{The multiplicity of 250-\mm~\textit{Herschel}\thanks{{\it Herschel} is an ESA space observatory with science instruments provided by European-led Principal Investigator consortia and with important participation from NASA.} sources in the COSMOS field}

\author[Scudder et al.] {Jillian M. Scudder$^{1}$\thanks{J.Scudder@sussex.ac.uk}, Seb Oliver$^{1}$, Peter D. Hurley$^{1}$, Matt Griffin$^{2}$,  Mark T. Sargent$^{1}$, \newauthor Douglas Scott$^3$, Lingyu Wang$^4$, Julie L. Wardlow$^5$ \\
$^1$ Astronomy Centre, Department of Physics \& Astronomy, University of Sussex, Brighton, BN1 9QH, England.\\
$^2$ Cardiff School of Physics and Astronomy, Cardiff University, Queens Buildings, The Parade, Cardiff CF24 3AA, Wales.\\
$^3$ Department of Physics and Astronomy, University of British Columbia, 6224 Agricultural Road, Vancouver, BC, V6T 1Z1 Canada \\
$^4$ SRON Netherlands Institute for Space Research, Landleven 12, 9747 AD, Groningen, The Netherlands.\\
$^5$ Centre for Extragalactic Astronomy, Department of Physics, Durham University, South Road, Durham DH1 3LE, England.\\}

\date{Accepted XXX. Received YYY; in original form ZZZ}

\pubyear{2015}

\begin{document}
\label{firstpage}
\pagerange{\pageref{firstpage}--\pageref{lastpage}}
\maketitle

\begin{abstract}
We investigate the multiplicity of extragalactic sources detected by the \textit{Herschel Space Observatory} in the COSMOS field. 
Using 3.6- and 24-\mm~catalogues, in conjunction with 250-\mm~data from \herschel, we seek to determine if a significant fraction of \herschel~sources are composed of multiple components emitting at 250 \mm. 
We use the {\sc xid+} code, using Bayesian inference methods to produce probability distributions of the possible contributions to the observed 250-\mm~flux for each potential component. The fraction of \herschel~flux assigned to the brightest component is highest for sources with total 250-\mm~fluxes $<45$ mJy; however, the flux in the brightest component is still highest in the brightest \herschel~sources. The faintest 250-\mm~sources (30--45 mJy) have the majority of their flux assigned to a single bright component; the second brightest component is typically significantly weaker, and contains the remainder of the 250-\mm~source flux. 
At the highest 250-\mm~fluxes (45--110 mJy), the brightest and second brightest components are assigned roughly equal fluxes, and together are insufficient to reach 100 per cent of the 250-\mm~source flux. 
This indicates that additional components are required, beyond the brightest two components, to reproduce the observed flux. 
95 per cent of the sources in our sample have a second component that contains more than 10 per cent of the total source flux. Particularly for the brightest \herschel~sources, assigning the total flux to a single source may overestimate the flux contributed by around $150$ per cent.
\end{abstract}


\section{Introduction}
\label{sec:intro}
As part of the drive to understand the evolution of galaxies over cosmic history, we look back at systems which existed at early times.  One of the more puzzling subsets of galaxies which have been identified is the population of very dusty, infra-red luminous, presumably highly star-forming galaxies at high redshifts. (For a review, see \citealt{Blain2002, Casey2014}.) 
As they contain a large amount of heated dust, these objects are typically very bright at far-infrared (FIR) or sub-mm wavelengths. Thus, the majority of the discoveries of these objects have come from FIR and/or sub-mm facilities.  
Attempts to identify optical counterparts to these sub-mm sources have historically been stymied by that same dust, as these objects are usually heavily extincted in the optical regime.  The `typical' galaxy of this type has so far been found to reside at redshifts of 1--3 \citep{Eales2000, Smail2002, Chapman2005}, though this is likely to partially depend on the selection wavelength \citep[e.g.,][]{Zavala2014, Bethermin2015}. 

This population of dusty, strongly star-forming sources at high redshift has drawn numerous comparisons to the local ultra-luminous infra-red galaxy (ULIRG) population \citep[e.g.,][]{Lonsdale2006}, as ULIRGs are the closest local analogues to the high redshift systems in terms of their unusually high star formation rates for their redshift.  ULIRGs, defined as having a total IR luminosity above $10^{12}$ L$_{\odot}
$\footnote{ULIRGs are defined by a fixed luminosity threshold; in the local universe this selects a rare subset of galaxies above the SFR-stellar mass relation, but at high redshifts the same luminosity threshold permits the inclusion of galaxies with less extreme SFR for their stellar mass \citep{Lin2007, Ellison2013}. The dusty, strongly star forming galaxies at high redshift are comparable to local ULIRGs in that they are both populations which are significantly offset from the SFR-stellar mass relation.}
, are similarly a rare subset of the local galaxy population, typified by unusually high star formation rates for their stellar mass, and contain a large volume of molecular gas {\citep{Combes2013}. 
If ULIRGs are truly local analogues of the high redshift systems, one might expect to find a substantial number of the high redshift galaxies are actually involved in some form of close galaxy-galaxy encounter, as suggested by several authors \citep[e.g.,][]{Tacconi2008, Younger2010, Zamojski2011, Wiklind2014}.  Unfortunately, it is difficult to unambiguously identify physical pairs at high redshift, and the fraction of the total samples which are ascribed to merger events therefore varies significantly. Both \citet{Tacconi2008} and \citet{Younger2010} find that their sources can all be explained with mergers, but sample only a few galaxies at the very brightest end.  \citet{Zamojski2011} has a much larger sample size, and estimates that their merger fraction should be 80 per cent, though this is extrapolated and could vary between 60 per cent and 91 per cent.
If the galaxies are not involved in an interaction to trigger their star formation, it remains unclear what secular mechanisms could drive a galaxy to such high star formation rates at early times in the universe.  

It is possible that the extremely high star formation rates implied for these FIR/sub-mm sources are boosted by the presence of a second, nearby source (either in projection or physically associated) which is also contributing to the FIR flux.
Single-dish observatories at the sub-mm and FIR wavelengths (whether space or ground based) typically have very large beam sizes (FWHM of order $\sim$ 20 arcsec, which at a redshift of 2 is $\sim$170 kpc), making it difficult to unambiguously identify a single optically detected galaxy as the source of the dust emission; the high space density of optical sources makes it likely that many potential counterparts will fall within the beam. \citet{Hughes1998} and \citet{Downes1999} both find 3-5 optical sources at the approximate location (i.e., within the positional error) of a sub-mm source (see also \citealt{Dunlop2004}).
If multiple optical components are contributing roughly equally to the FIR flux, then a single FIR source may well be comprised of multiple, potentially independent physical sources.
It is difficult to determine whether the optical counterparts are indeed contributing to the sub-mm flux, and often a single counterpart is assigned responsibility for the entirety of the sub-mm flux (typically the counterpart with the smallest projected separation), though in some cases other methods, such as a Bayes factor, have been used \citep[e.g.,][]{Roseboom2009}.
The issue of multiple possible sources will always be present in some fraction of the data, as there will be line of sight projection effects which will layer unrelated foreground or background sources in the same region of sky as the intrinsic FIR source. 

To overcome the challenge of identifying the correct source of the FIR flux, several methodologies have been developed.  One approach is to use data from other wavelengths, such as the radio, where higher resolution imaging of the same area of the sky can be obtained, making use of the strong radio-FIR correlation \citep[e.g.,][]{Helou1985}.  Radio observations provided some of the earliest hints that a significant fraction of the galaxy population might be composed of blended components \citep[e.g.,][]{Ivison1998, Ivison2000, Smail2000, Ivison2002}.  
More recently, it has become possible to make use of interferometric facilities such as the Sub-Millimetre Array \citep[SMA; e.g.,][]{Younger2007,Younger2008, Wang2011c}, the Plateau de Bure Interferometer \citep[PdBI; e.g.,][]{Smolcic2012},  or the Atacama Large Millimetre/sub-millimetre Array \citep[ALMA; e.g.,][]{Hodge2013}, which have the power to observe some of these systems with sub-arcsecond resolution.

Recent ALMA results have added to a growing pool of bright FIR-selected and sub-mm sources which resolve into multiple components.  Whereas previous work had focused on deep optical imaging or small samples \citep[e.g.,][]{Younger2010, Wang2011c, Chen2013b} to find potential counterparts, ALMA can observe in the FIR with high resolution, avoiding complications which arise when comparing with other wavelengths \citep[e.g.,][]{Hodge2013}.  However, the fraction of sources which divide into multiple components varies substantially between studies. Multiplicity fractions (defined as the fraction of total sources which resolve into multiple components) appear to range between 10 per cent \citep{Chen2013b} and 50 per cent \citep{Hodge2013}, with other studies finding values within that range \citep[e.g.,][measuring 25 per cent and $>$ 20 per cent fractions respectively]{Barger2012, Hezaveh2013}\footnote{For the works cited here, the beam size is not substantially different between studies, varying between 3 arcsec \citep{Chen2013b} and 1.5 arcsec \citep{Hodge2013}, but may play a role when comparing to other analyses.}.
This variation is likely due in part to different sample selection between studies, which select their targets based on flux measurements based anywhere between 450-\mm~ \citep[e.g.,][]{Chen2013b} and 1.2 mm  \citep[e.g.,][]{Hezaveh2013}, though largely the systems are selected close to 850-\mm~\citep{Hodge2013, Karim2013, Simpson2015}.  Another possible source of variation is the indication that the multiplicity fraction may be dependent on the FIR flux observed, with brighter sources more likely to have multiple components \citep{Karim2013}. While ALMA has the advantage of directly observing with high resolution in the FIR, it is limited by sky coverage.  
In this work, we will make use of the availability of multi-wavelength coverage of large areas of the sky, which allows us to make use of higher resolution data to determine how many FIR sources divide into multiple flux-emitting components.

If a significant fraction of the FIR sources are in fact multiple components, then \textit{the individual galaxies are much less extreme systems than previously thought}, regardless of whether or not the component galaxies are physically interacting with each other.  In order to develop a better understanding of how the population of star forming galaxies has evolved over time, we must determine the exact nature of these seemingly extreme systems.
We wish to undertake a comprehensive survey of the multiplicity of these dusty star forming galaxies using a new approach: Bayesian statistical analysis of multi-wavelength coverage of a given field.  This study will be the first to undertake such a statistical survey of bright \herschel~sources.    

In Section \ref{sec:data}, we describe the data used for this study, including a brief description of the multi-wavelength catalogues.  In Section \ref{sec:sample}, we describe the selection criteria used for the data, and in Section \ref{sec:analysis} we describe the methods used to analyse this data.  In Section \ref{sec:discussion}, we discuss the implications of our results in the context of previous works, and in Section \ref{sec:conclusions} we summarise our conclusions.

\section{The Data}
\label{sec:data}
We base our study on data from the \textit{Herschel Space Observatory} \citep{Pilbratt2010}.  \herschel~was equipped with two continuum imaging instruments, the Photodetector Array Camera and Spectrometer \citep[PACS;][]{Poglitsch2010}, and the Spectral and Photometric Imaging Receiver \citep[SPIRE;][]{Griffin2010}.  Both instruments observed a number of legacy fields as part of the  \textit{Herschel} Multi-Tiered Extragalactic Survey \citep[HerMES\footnote{\url{hermes.sussex.ac.uk}};][]{Oliver2012}, which provided FIR coverage at 100 and 160 \mm~in PACS, and at 250, 350, and 500 \mm~from SPIRE.

A new EU funded project, the \herschel~Extragalactic Legacy Project (HELP\footnote{\url{http://herschel.sussex.ac.uk/}}; e.g., \citealt{Vaccari2015}), aims to collect and reprocess data products across multiple wavelengths for all regions contained within the \herschel~survey fields, including fields in surveys not contained in HerMES, such as \herschel-Astrophysical Terahertz Large Area Survey \citep[H-ATLAS;][]{Eales2010} and Herschel Stripe 82 Survey \citep[HerS;][]{Viero2014}.  The main goal of HELP is to produce homogeneously calibrated, value-added catalogues covering roughly 1,000 deg$^2$, which matches individual sources across wavelengths, allowing for statistical, multi-wavelength studies of the local-to-intermediate redshift galaxy population.  In this work, we benefit from the multi-wavelength ancillary catalogues already compiled by HELP.
For the current work we limit our study to the COSMOS field \citep{Scoville2007}, a field observed by both PACS and SPIRE instruments as part of the PACS Evolutionary Probe (PEP) program \citep{Lutz2011} and HerMES \citep{Oliver2012}, and which has excellent multi-wavelength coverage over the 2 deg$^2$ field.  We focus on the 250-\mm~data catalogue from the SPIRE instrument, as it has the highest SPIRE resolution, and of the three SPIRE bands will be the least affected by blending.  Our \herschel~source list is produced by the {\sc starfinder} software and is a completely blind catalogue \citep{Smith2012a, Wang2014a}. The \herschel~maps are generated by the {\sc smap/shim} software and are a standard HerMES Data Release 2 product \citep{Levenson2010}.

Traditionally, studies of multiplicity within the \textit{Herschel} catalogues have been tackled using the positions of 24-\mm~sources from the \textit{Spitzer Space Telescope} \citep{Werner2004} as more precise indicators of the position of likely sources within a 250-\mm~data set \citep[e.g.,][]{Roseboom2010, Brisbin2010}.  A similar method was used by \citet{Smith2011a}, employing positional information to assign H-ATLAS source flux to a counterpart in the Sloan Digital Sky Survey Data Release 7 \citep{Abazajian2009}.  For this study, we use the positions of sources from the 24-\mm~\textit{Spitzer} Multiband Imaging Photometer (MIPS) catalogue \citep{LeFloch2009} and the 3.6-\mm~\textit{Spitzer} Infrared Array Camera (IRAC) catalogue, which were produced as part of S-COSMOS \citep{Sanders2007}. The 24-\mm~catalogue was extracted from the maps using the SExtractor software \citep{Bertin1996} to identify sources, with fluxes determined through the use of {\sc daophot} \citep{Stetson1987}. The 24-\mm~catalogue is estimated to be over 90 per cent complete above a flux of 80 $\mu$Jy \citep{LeFloch2009}; our catalogue is flux limited to contain only sources above 150 $\mu$Jy, so our sample should be complete. 

The 3.6-\mm~catalogue contains $5\sigma$ detections down to a flux of 0.9 $\mu$Jy, and is also extracted from the map using the SExtractor software \citep{Bertin1996, Sanders2007}. 
The 3.6-\mm~data provide a much higher number of potential counterparts than the 24-\mm~data, since it is much higher resolution (1.7 arcsec at 3.6 \mm~vs 6 arcsec at 24 \mm).  By using the 3.6-\mm~data, our resolution information has improved by a factor of 10, relative to the 250-\mm~data.
The 3.6- and 24-\mm~sources will be used to help us probe the question of how frequently FIR sources divide into multiple components; by using the higher resolution information from shorter wavelength data, we can use the positions of these sources as prior information for our analysis of the SPIRE photometry.

\section{Sample Selection}
\label{sec:sample}
We wish to select a sample of \herschel~sources that have higher resolution counterparts in order to examine the frequency with which they divide into multiple components. We therefore identify a sample of 250-\mm-detected sources that have at least one 3.6-\mm~source and at least one 24-\mm~source within the size of the \herschel~beam at 250 \mm~(18.1 arcsec; \citealt{Griffin2010}).

We begin our sample selection by including in our master catalogues only reasonably secure detections.  We include all 250-\mm~sources within the COSMOS field above a flux limit of 30~mJy, which corresponds to approximately $5\sigma$ above the confusion noise limit in the \textit{Herschel} observations \citep{Nguyen2010, Oliver2012}.  Confusion noise dominates over all other sources of noise in the SPIRE COSMOS data; see \citet{Nguyen2010}.
As a first quality control measure on both the 24-\mm~catalogue \citep{LeFloch2009} and the 3.6-\mm~catalogue, we require that the flux values recorded are above zero, since negative flux values are unphysical.  For the 3.6-\mm~catalogue \citep{Sanders2007}, we additionally require that the source flux flag is set to 0, which indicates that the source flux is not contaminated by a nearby bright source in the IRAC catalogue\footnote{In the IRAC catalogue, flag = $0$ is shortened to ``good data''.}. 

For each 250-\mm~source in the COSMOS field, we search for those that have at least one 24-\mm~source that passes our quality control measures within the 18.1-arcsec full width half maximum of \herschel~at 250 \mm. We require a 24-\mm~detection for our sample, as many previous works (including standard HerMES products) have required 24-\mm~detections at the locations of a \herschel~source to ensure that their \herschel~detections are reliable astrophysical sources. For consistency with these methodologies, we impose the same criterion on our sample.  However, the 24-\mm~data is of limited spatial resolution, so to increase the resolution information, we also require a 3.6-\mm~detection.
We use the same procedure to select out all 250-\mm~sources that have at least one 3.6-\mm~source within the 250-\mm~beam.  We then cross-match the 3.6-\mm~matches with the 24-\mm~matches, to select those 250-\mm~sources that have potential components in both the 24-\mm~and 3.6-\mm~catalogues.  This cross-matching marks the selection of the final sample of 250-\mm~sources.

For each of our final 250-\mm~sources, if a 24-\mm~source is within 6.0 arcsec of a 3.6-\mm~source matched to the same 250-\mm~source (6 arcsec is the resolution of \textit{Spitzer} at 24 \mm), we remove the 24-\mm~location, making the assumption that the 3.6-\mm~source will be a more accurate position for the same physical object. This process does not exclude single-counterpart \herschel~sources from our sample, as we do not require more than one possible counterpart per detected \herschel~source. In some cases the 24-\mm~object is farther than 6.0 arcsec away from the nearest 3.6-\mm~source. In these cases, we keep both the 3.6- and the 24-\mm~object positions for our list of potential contributing components for the 250-\mm~source\footnote{Within our sample, this is true in 25/360 of our 250-\mm~sources.}.

Our final sample consists of 360 \textit{Herschel} 250-\mm~sources within the COSMOS field, each having at least one possible contributing source at 24 \textit{and} at 3.6 \mm. The limiting criterion is requiring a 24-\mm~source within the 250-\mm~beam, which restricts us to 14 per cent of the total 250-\mm~catalogue in this field. However, this sample represents 98 per cent of the 250-\mm~sources with 24-\mm~detections within the \herschel~beam; the inclusion of the 3.6-\mm~criterion only reduces the sample by 2 per cent. Although we did not explicitly require that each 250-\mm~source have more than one potential component at 24 and 3.6 \mm, the fewest number of potential components matched to each \herschel~source is 3. The maximum number of potential components within the 250-\mm~beam is 26; the median number of candidate sources is 14 per 250-\mm~source.

\subsection{Using {\textsc{xid+}} to assign component fluxes}
With many potential contributing sources per 250-\mm~source,  we must find a way to assign flux to each individual component if we are to investigate the properties of the underlying components.
To do so, we make use of the {\textsc{xid+}} software \citep{Hurley2016}, a prior-driven source-extraction tool, and a significant improvement upon the previous {\sc{xid}} software, also known as {\sc{desphot}} \citep{Roseboom2010, Wang2014}.  This new software's methodology and validation are fully described in\citet{Hurley2016}, so we refer the reader to that work for a complete explanation of the methods, with a brief explanation here.  

The {\sc{xid+}} algorithm uses the 250-, 350-, and 500-\mm~maps from \herschel, along with the positional information of known sources to find the most likely distribution of flux among these sources. {\textsc{xid+}} assigns the appropriate \textit{Herschel} flux to each of the sources at the known input positions.  {\textsc{xid+}} places a positive prior on the fluxes; in other words it cannot assign negative flux to a source to improve the fit, as this has no physical meaning for observations.  The code has been tested on synthetic maps of the COSMOS field used here and is found to recover injected source fluxes accurately and with a reliable error estimate.  The previous code {\sc desphot} is found to less reliably recover the true flux of injected sources, and also systematically underestimates the errors on those measurements \citep{Hurley2016}. This difference is largely due to {\sc desphot}'s use of a maximum likelihood algorithm which gradually switches on sources to increase the goodness of fit; sources which do not improve the fit quality are assigned a flux of 0 mJy. We consider XID+ to be a considerable improvement over the methods used in {\sc desphot}.  

For this work, since we have already selected a subsample of the 250-\mm~catalogue, we do not need to run {\textsc{xid+}} on the full COSMOS map, which would require us to simultaneously fit several tens of thousands of sources.  Instead, we select a 180 arcsec by 180 arcsec box surrounding each 250-\mm~source\footnote{The exact size of this box is somewhat arbitrary, but this ensures that the entire 250-\mm~source, along with a buffer region surrounding it, are fit by {\textsc{xid+}} for each 250-\mm~source.}.
We then find all of the 3.6-\mm~sources within that field (typically several hundred in number), in addition to those identified as potential counterparts, in order to ensure a good fit to the entire region, which will ensure that we do not neglect to assign flux to other sources near our 250-\mm~source of interest. We neglect the 24-\mm~positions here as the typical field contains no additional 24-\mm~sources which are not coincident with a 3.6-\mm~source, so the inclusion of the 24-\mm~sources is unlikely to improve the fit.
This full list is then used as the `known sources' for the positional prior information required by {\textsc{xid+}}.  {\textsc{xid+}} is then run on each of the 360 unique fields surrounding our 250-\mm~sources, and the resulting flux probability distributions are stored for all of the fitted sources, not just those which fall within the beam of the strong 250-\mm~source.  However, for the analysis which follows, we investigate only those potential counterparts which were flagged as within 18.1 arcsec of the \herschel~source.

{\textsc{xid+}} produces a full Bayesian statistical posterior probability distribution through MCMC sampling, which allows strongly correlated sources to be identified.  In cases where more than one component is likely to be producing FIR flux, the posterior allows us to examine all of the possible permutations of flux assignment between these components.  The full posterior also provides accurate error estimates on the flux estimates.  In instances where the final probability distribution is Gaussian, or nearly Gaussian, taking the median, 16th and 84th percentiles of the probability distribution will provide an accurate summary of the typical flux and its error.  Where the fluxes assigned are very low (they are constrained to remain above zero) the median will tend to be biased towards slightly higher values than are truly typical (i.e., the calculated median may not coincide with the mode of the distribution). 
The raw probability distributions can be viewed through a corner plot \citep[][Figure \ref{fig:triangle}]{triangle-py}.  

\begin{figure*}
   \centering
   \includegraphics[width=500px]{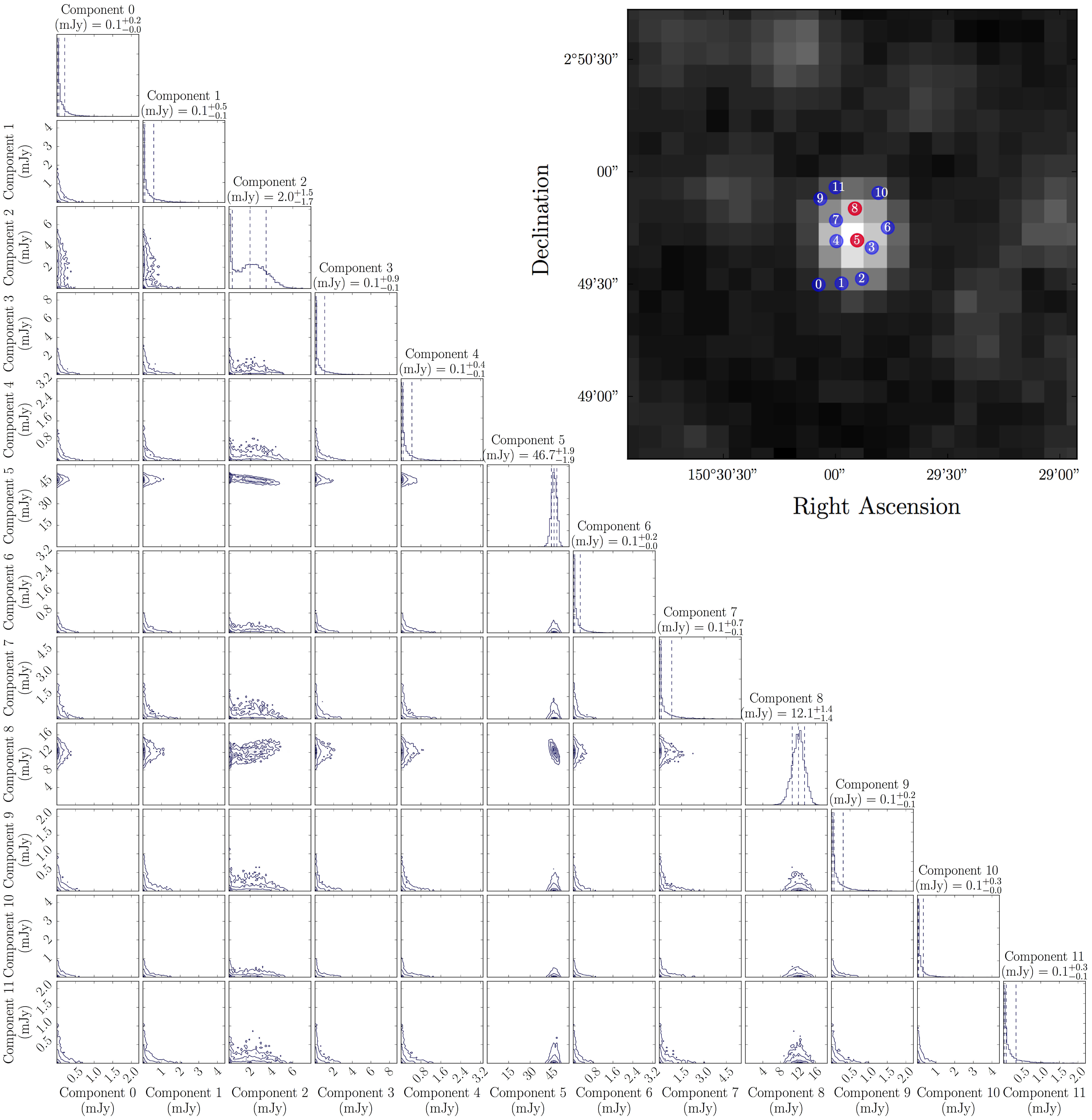}
   \caption{A triangle plot of the correlations between the 12 potential components for a single 250-\mm~source (ID 10009). Histograms along the top diagonal edge indicate the probability distribution of flux for the component in that column. Above each histogram, the 50th percentile value for the component is labelled, plus/minus the distance to 16th and 84th percentiles. Vertical dashed lines indicate the 16th, median, and 84th percentiles (from left to right in each panel). The interior panels give the contours of correlation between pairs of components. In the top right corner, the 250-\mm~map is shown in grayscale, with the positions of the known potential counterparts shown as coloured circles. Each potential component is labelled as in the triangle plot. We can see that only two sources are assigned significant flux, with the majority of the flux assigned to Source 5. We note that the axis ranges are not consistent across all components, though all axes begin at 0 as their lower limit. With the exception of components 5 \& 8, and to a lesser extent component 2, all other sources have distributions peaking very close to 0 flux.  Sources 5 and 8 are therefore marked in red in the inset figure.}
   \label{fig:triangle}
  \end{figure*}
Figure \ref{fig:triangle} shows the probability distributions for each of the 12 components of the 250-\mm~source in a typical field.  The majority of the components are given fluxes that are very close to zero, with the bulk of the flux assigned to component 5 and component 8.  Of the two sources with significant flux, component 5 dominates the flux budget, with a median flux of 46.7 mJy, and the 16th and 84th percentiles falling at 44.8 and 48.6 mJy respectively. 16 and 84th percentiles for component 8's flux set a range between 10.7 and 13.5 mJy, with component 2 receiving a median of 2.0 mJy, with the 84th percentile setting an upper limit of 3.5 mJy of flux.  The contours which populate the off-diagonal show the correlations between components.  There is a slight correlation between component 5 and component 8, but the probability distributions for these two components are relatively narrow, indicating that the possible error due to degeneracy in the solution is minimal.
In the case where two components are completely degenerate, the probability between sources will appear as a tightly correlated diagonal across a wide range of possible fluxes.  This poor constraint on parameter space would be reflected in the width between the 16th and 84th percentile values being very large.
We have inspected all the corner visualisations for the 360 fields in our sample, and find that flux is typically assigned to 2--4 components out of a median number of 14 potential components\footnote{The full set of triangle plots is available for public download at the following http URL: \url{http://jmscudder.github.io/XID-figures}}.

We note that {\sc xid+} uses a prior on the fluxes which is flat in parameter space.  While this prior has the effect of preferring solutions with a larger number of lower flux objects relative to higher flux objects, this is a physically motivated prior, as very bright sources should be rarer objects than objects which are intrinsically fainter.  We also note that {\sc xid+} is limited to assigning fluxes to objects which are provided as known sources.  However, if the known source list was very incomplete, {\sc xid+} would find a poorer solution to the map, which would be reflected in a broader width of the posterior distribution.  We account for the uncertainty in the posterior distribution throughout this work, and do not believe that the 3.6-\mm~catalogue is a significantly incomplete representation of the FIR-emitting population \citep{Duivenvoorden2016}; these effects should not bias our results.

{\textsc{xid+}} also includes statistical methods to test for convergence, allowing us to be certain that the parameter space has been appropriately explored.  We require that for each chain, the individual runs have completely converged.  Following the guidelines presented in \citet{Hurley2016}, we constrain \^R\footnote{\^R is the potential scale variation as estimated from a measurement of the posterior variance within chains of the MCMC method, and measures how well each run has found the global minimum.} to be $< 1.2$, and n$_{\mathrm{eff}}$\footnote{n$_{\mathrm{eff}}$ is the effective number of samples, and checks if the global minimum has been well traversed.  Both convergence metrics are fully explained in \citep{Hurley2016}. The limits used here are found to ensure that the code has converged.} $> 40$.  For the work done here, we use 4 MCMC chains of 2000 iterations each, which is sufficient to meet the convergence criteria for 99.5 per cent of all runs.

\section{Analysis}
\label{sec:analysis}
Our ultimate goal in analysing the results of {\textsc{xid+}} is to determine what fraction of these 250-\mm~sources have strong multiple components, and if there are systematic differences within the population we have selected which depend on the brightness of the 250-\mm~source.
To investigate whether we have a fair base of comparison within the sample, we first wish to determine if there are systematic variations in the number of potential counterparts surrounding the selected 250-\mm~sources as a function of 250-\mm~flux. 
If there is a systematic change in the number of components, then any properties which are plotted as a function of the fraction of total flux or as a fraction of the number of components will also be systematically biased. 
If we find that there is no strong systematic shift with total 250-\mm~source flux, then the raw number of potential components will not be a source of bias in our analysis of trends with total 250-\mm~flux \footnote{We note that a uniform underrepresentation of potential components across all 250-\mm~sources will still have an effect on, e.g., number counts, where the absolute number of sources is important.  For the current work, we are not sensitive to this issue.}.
In Figure \ref{fig:cumul_numsource}, we plot the cumulative distribution of the number of potential components found per 250-\mm~source, divided into bins of total flux across all potential components.  Bins here are defined such that each bin has an equal fraction (20 per cent) of the total sample. We have also binned the data evenly in total flux, and the results are broadly unchanged, but at high 250-\mm~source flux, bins suffer from low number statistics. 

 We note that the lowest flux limit in Figure \ref{fig:cumul_numsource} is slightly below 30 mJy, which was our original flux limit described in \S \ref{sec:sample}.  Here, and throughout this work, we use a total 250-\mm~calculated from the output of {\sc xid+}. For each iteration of {\sc xid+}, we sum the fluxes of all components identified as within 18.1 arcsec.  We then use the median of the distribution of summed fluxes as our total flux throughout this analysis. In the case of the lowest fluxes plotted here, 4 mJy of flux has been assigned to a component outside of the beam of the 250-\mm~source, lowering the total flux below our nominal 30 mJy threshold.

To determine if the bins are statistically consistent with each other, we run a Kolgolmorov-Smirnoff (KS) test on the distributions. 
A KS test finds the maximum vertical separation of the cumulative distributions of two populations to determine if the two populations are consistent with being drawn from the same parent population.  If the maximum distance is too large, the distributions are inconsistent. Each of the five bins is compared with the remaining four in turn. We find that the KS test indicates that all bins are consistent with being drawn from the same parent sample as all other bins\footnote{Our most discrepant value (between the lowest and highest 250-\mm~flux bins) give p-values of 0.144, or less than $1.5\sigma$.}. The 250-\mm~sources are therefore not populated with a significantly different number of sources in the field as a function of total flux.  
If the 250-\mm~sources have consistent numbers of potential components as a function of total flux, we should be able to make robust relative comparisons as a function of total flux for other properties of the components.

\begin{figure}
   \centering
   \includegraphics[width=250px]{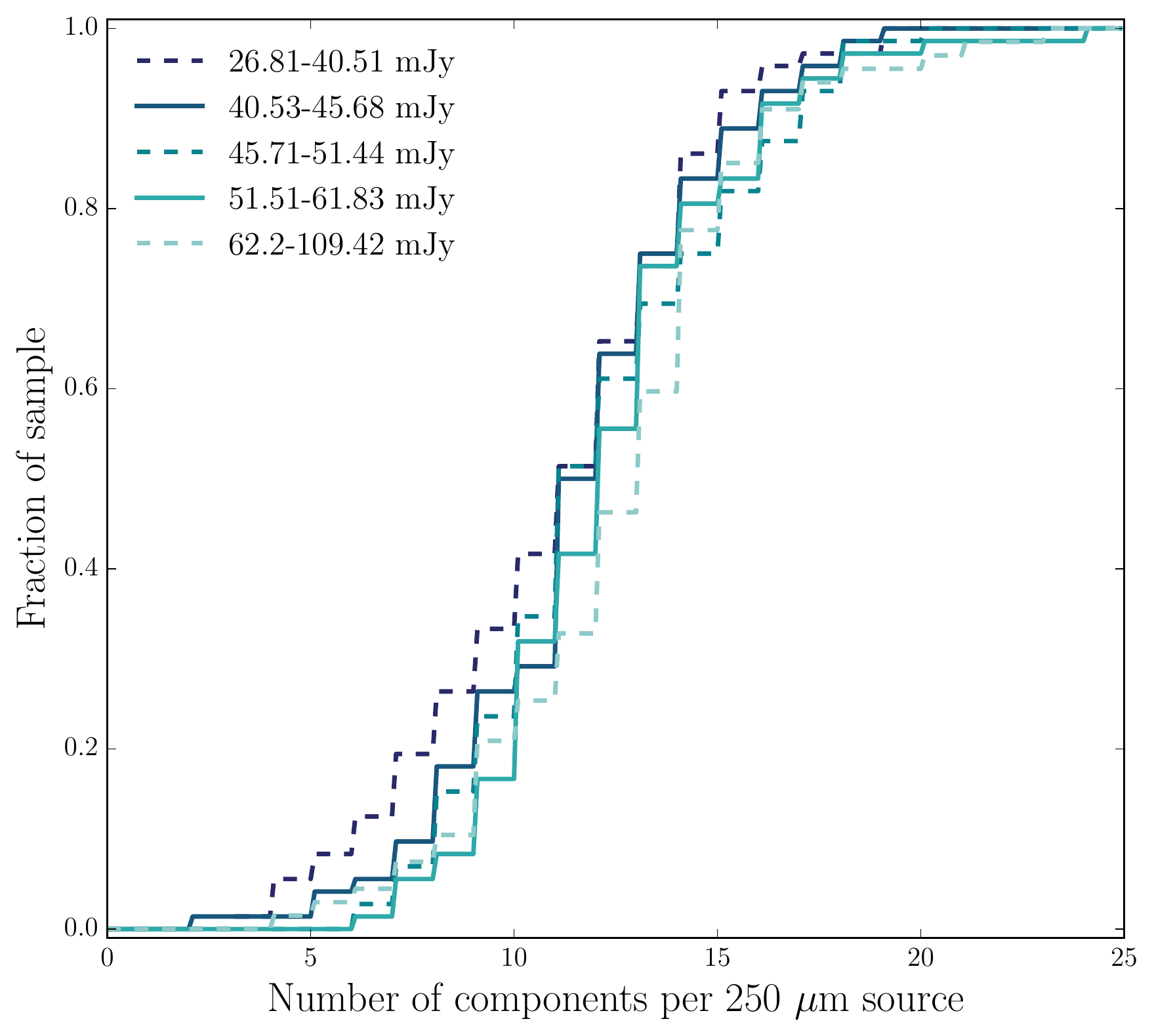}
   \caption{The cumulative distribution of the number of all potential components per 250-\mm~source, binned as a function of total 250-\mm~source flux. Each bin contains an equal fraction of the total sample.  Line style alternates between dashed and solid for added visual clarity.  A KS-test indicates that all flux bins are consistent with being drawn from the same parent population as all other bins.}
   \label{fig:cumul_numsource}
  \end{figure}

In Figure \ref{fig:unbinned} we plot the fraction of the total 250-\mm~flux assigned to each component of our full sample.  Each data point indicates the median of the posterior distribution of possible fluxes.
We plot the data as a density histogram, where colour indicates the log of the number of components within that bin, indicated by the colourbar on the far right of the figure.  Note that, as Figure \ref{fig:triangle} implied, there are a large number of components with near-zero fluxes, which results in a concentration of potential components along the x-axis (visible as the dark blue stripe).  On the right hand panel of Figure \ref{fig:unbinned}, we show the 1D histogram of component flux fractions (with the horizontal axis plotted in log), collapsed along the total 250-\mm~source flux axis.  This panel emphasises the abundance of components assigned very little flux.  Curiously, we note that the upper right hand corner of this diagram appears unpopulated by data points.  
We also note that in our density histogram, the low 250-\mm~source flux end appears to have a much higher concentration of points with zero flux than is present at the high 250-\mm~source flux end.  
To verify that this is not simply due to the flux distribution of 250-\mm~sources, we divide the sample into the same set of equally populated bins as Figure \ref{fig:cumul_numsource} and compare the distribution of flux fractions in each bin to all other bins with a KS test. For the full sample, the KS test between the three lowest total flux bins and the highest total flux bin exclude the null hypothesis of being drawn from the same parent population at much more than $5\sigma$. The second highest flux bin is also inconsistent with the highest flux bin at $> 4 \sigma$. This indicates that the distributions of flux fraction assigned to components are statistically distinct from each other as a function of total 250-\mm~flux.
\textit{A priori}, this might lead us to expect that a larger fraction of the low 250-\mm~flux components are being assigned very low flux values, and that at high 250-\mm~flux values, significant flux is being divided among more components.  We investigate this further in the following sections.

\begin{figure}
   \centering
   \includegraphics[width=250px]{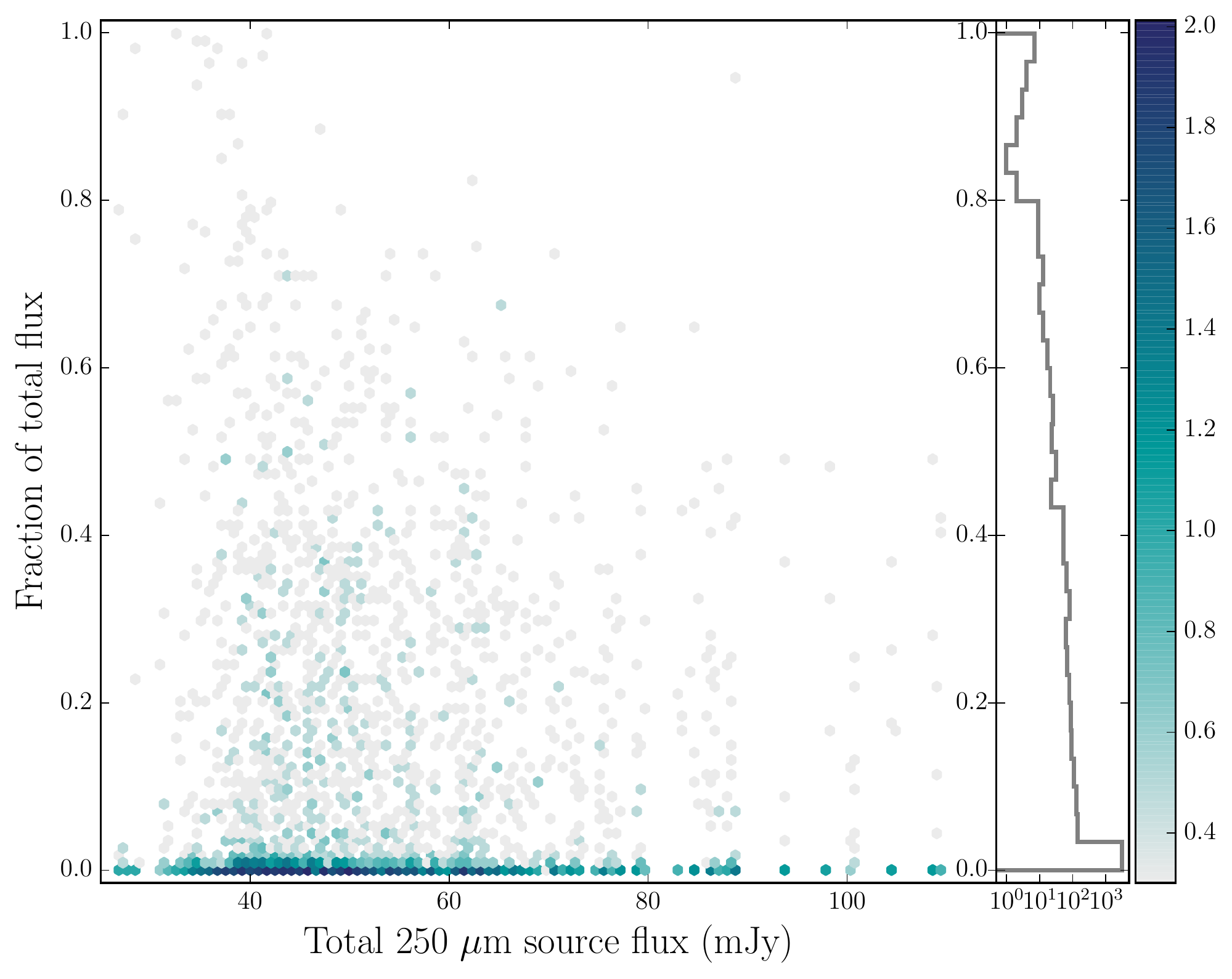}
   \caption{Left panel: the unbinned data points for all potential components in our sample, plotted as a density histogram.  We plot the fraction of the total 250-\mm~source flux assigned to each potential component, as a function of the total 250-\mm~flux. Colour indicates the log of the number of points contained within that bin, as indicated by the colourbar on the far right.  Points close to 1 on the y-axis indicate that they were assigned almost 100 per cent of the total flux, meaning that they are the only source assigned significant flux. For high total flux 250-\mm~sources, there is a deficit of components at large flux fractions (i.e., high flux objects tend to have that flux divided amongst more components, explored in more detail in text). The right hand panel shows a histogram of the distribution of flux fractions across all 250-\mm~source fluxes.  The x-axis in the right panel is in log space.}
   \label{fig:unbinned}
  \end{figure}
  
\subsection{The brightest components}
As our sample has a large number of components which have been assigned extremely low flux, it may be that components assigned low flux values are swamping any interesting trends within the fraction of the sample which has been assigned \textit{significant} flux.  To investigate, we use the brightest component from each of the 360 250-\mm~sources in our sample.

It is possible that some of the 250-\mm~sources have two (or more) nearly equal brightness components.  In this case, identifying a physical component as the brightest component is potentially unwise, as the posterior distribution may indicate that it is equally likely for the second component to be the brightest.  
The same physical component will not necessarily always be the brightest in all of the iterations.
To circumvent this problem, we select the brightest component on a sample by sample basis.  We therefore accumulate a distribution of brightest component fluxes for each of the 2000 iterations.  We use the median of this distribution as the `brightest component flux' for that 250-\mm~source.  We place these brightest components into equally populated bins of 250-\mm~flux, and plot them as a function of 250-\mm~source flux in Figure \ref{fig:bin_maxfraction}.  The horizontal error bars in Figure \ref{fig:bin_maxfraction} reflect the width of each bin.  Below each point we label the number of 250-\mm~sources contained within that bin.

To make an estimate of the error on the median flux of the brightest components in each bin, we calculate the errors in two ways.  The first takes into account the full posterior distribution of the brightest component flux measurement itself.  To calculate the error, we resample the posterior of each brightest component contained within a given 250-\mm~bin.  If the brightest component flux varies strongly across iterations, we will capture this variation with the resampling.  For each sampling of the distribution of each brightest component, we recalculate the median of that bin.  This is repeated for all 2000 iterations, to fully sample the distribution of each brightest component.  The standard deviation on the distribution of resampled median binned points is plotted as the pale grey vertical error bars in Figure \ref{fig:bin_maxfraction}.  This error bar reflects not only the dispersion within the population contained within that bin, but also the possible measurement error due to variations in how well constrained the fluxes are.

However, each of the brightest components within the bin is given a completely independent flux measurement, as they are fitted by {\sc xid+} separately.  We do not expect the measurement error to have any systematic effect on the medians, so we plot the above described error bars divided by the square root of the number of brightest components contained within the bin (shown in Figure~\ref{fig:bin_maxfraction} as the number plotted below the error bars).  This $\sqrt{n}$ error bar is plotted in black.  Typically, the $\sqrt{n}$ set of error bars is approximately the size of the plotted median point or smaller.
While we plot both error bars for the remainder of this work, we expect that for trends within the sample, the $\sqrt{n}$ error bars are a better reflection of our expectation of the uncertainty in the median itself, and refer to those error bars for the remainder of this work.

In the binned values, it is clear that the median flux fraction in the brightest component increases at the low total 250-\mm~flux end of the diagram.  To determine if this is statistically significant, we run a Spearman rank correlation test on the \textit{unbinned} data. This tests if two properties are correlated by checking if both quantities increase with each other, but has the advantage of not requiring that this correlation be linear, as (e.g.) the Pearson test does. By running the correlation on the unbinned data we are also insensitive to our choice of bin.
The Spearman Rank correlation test returns a p-value of $1.86\times10^{-12}$, which excludes the null hypothesis of no correlation at $>6\sigma$. The Spearman rank test also returns a correlation coefficient of $-0.36$.
\begin{figure}
   \centering
   \includegraphics[width=250px]{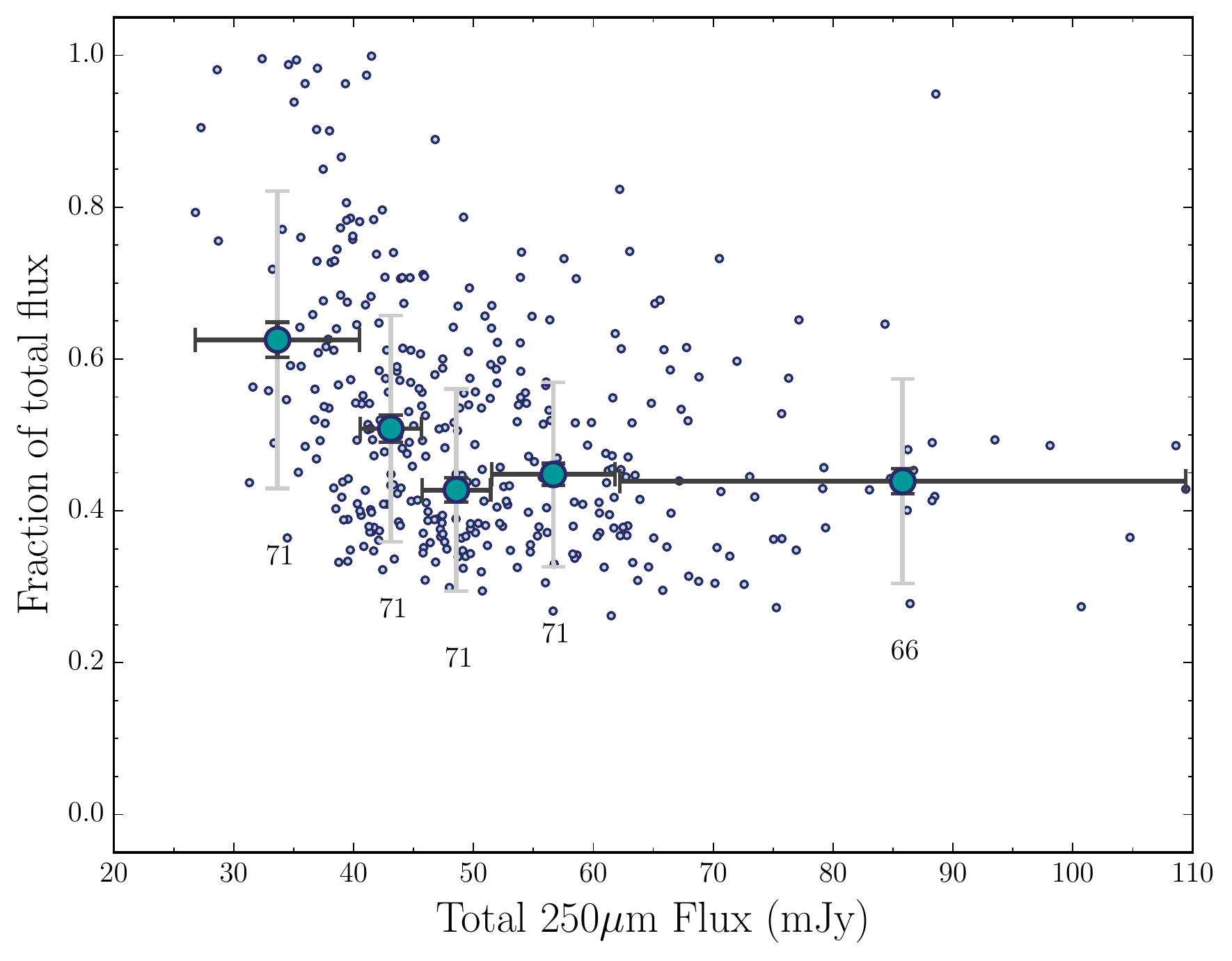}
   \caption{For each 250-\mm~source, we select the component with the highest assigned flux fraction.  We plot these brightest components as a function of the total 250-\mm~flux in small background points. Horizontal error bars on the larger, binned points indicate the width of the bin, and the numbers below each point indicate the number of brightest components within each bin. Pale vertical error bars show the standard deviation on the distribution of all possible medians, including both the flux uncertainty and the dispersion of brightest component contained within the bin. Black vertical error bars (typically of comparable size to the data point) are calculated by dividing the pale error bars by $\sqrt{n}$, applicable as each data point is an independent measurement and measurement error should not introduce a systematic bias. We refer to the black $\sqrt{n}$ error bars for the remainder of this work. The brightest components show a trend towards higher fractions of 250-\mm~flux as total 250-\mm~flux decreases.  A Spearman Rank correlation test on the unbinned data gives a p-value of $1.86\times10^{-12}$, which excludes the null hypothesis of no correlation at $>6\sigma$.}
   \label{fig:bin_maxfraction}
\end{figure}

We also investigate the raw flux assigned to each component. The exact relationship between brightest component flux and 250-\mm~flux can be inferred from Figure \ref{fig:bin_maxfraction}. As the flux fraction dependence on 250-\mm~flux is not linear, we plot the dependence of brightest component flux on 250-\mm~flux explicitly in Figure \ref{fig:bin_maxflux} for clarity. This figure is otherwise constructed identically to Figure \ref{fig:bin_maxfraction}.  Here we see a strong increase in the amount of flux assigned to the brightest component as a function of total flux, with the Spearman Rank test applied to the unbinned data indicating a p-value of $1.08\times10^{-13}$, an exclusion of the null hypothesis (no correlation) at $>7\sigma$.  The Spearman Rank correlation test indicates a correlation coefficient of 0.37.

\begin{figure}
   \centering
   \includegraphics[width=250px]{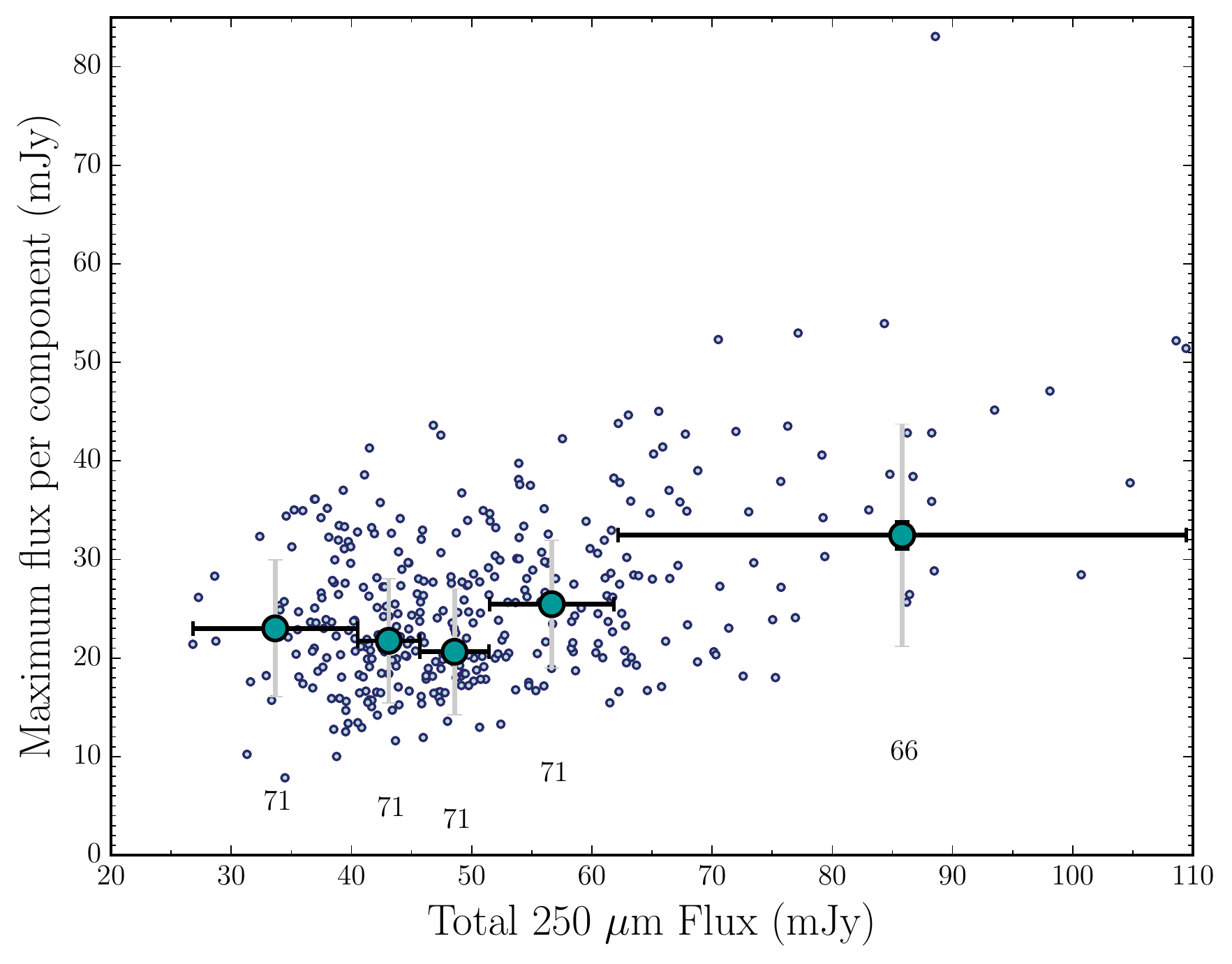}
   \caption{The same as Figure \ref{fig:bin_maxfraction}, but now plotting the flux value assigned to the brightest component instead of the fraction of the total flux. The brightest components at high 250-\mm~fluxes are still brighter than the low 250-\mm~flux components. A Spearman Rank correlation test on the unbinned data gives a p-value of $1.08\times10^{-13}$, indicating an exclusion of the null hypothesis of no correlation at $>7\sigma$. } 
   \label{fig:bin_maxflux}
\end{figure}

Together with Figure \ref{fig:bin_maxfraction}, Figure \ref{fig:bin_maxflux} shows that while the brightest components at lower total 250-\mm~flux often have a single component with a significant fraction of the total flux of the 250-\mm~source, those components are still assigned significantly lower fluxes than the components at high total 250-\mm~fluxes, which are sharing flux amongst multiple objects. While bright 250-\mm~sources divide their flux into multiple objects, the brightest of the high flux components are still brighter than the lower 250-\mm~flux brightest components\footnote{We note that our binned points at fluxes lower than 50 mJy have a poor positive correlation.  However, this is due to our choice of bins, and disappears if we use bins which are equal in width. The full population (plotted as background small points) does not have a strong break at 50 mJy.}.

\subsection{Brightest component vs. 2nd brightest component}
\label{sec:2ndbrightest}
As an additional check into the behaviour of these brightest components, we investigate the relationship between the fraction of flux assigned to the brightest component and the fraction assigned to the second brightest component. We identify the second brightest component in an identical fashion to identifying the brightest component. In Figure \ref{fig:unbin_1to2}, the ratio of the brightest component's flux to the second brightest component's flux is plotted in log against the total 250-\mm~source flux for each 250-\mm~source.

\begin{figure}
   \centering
   \includegraphics[width=250px]{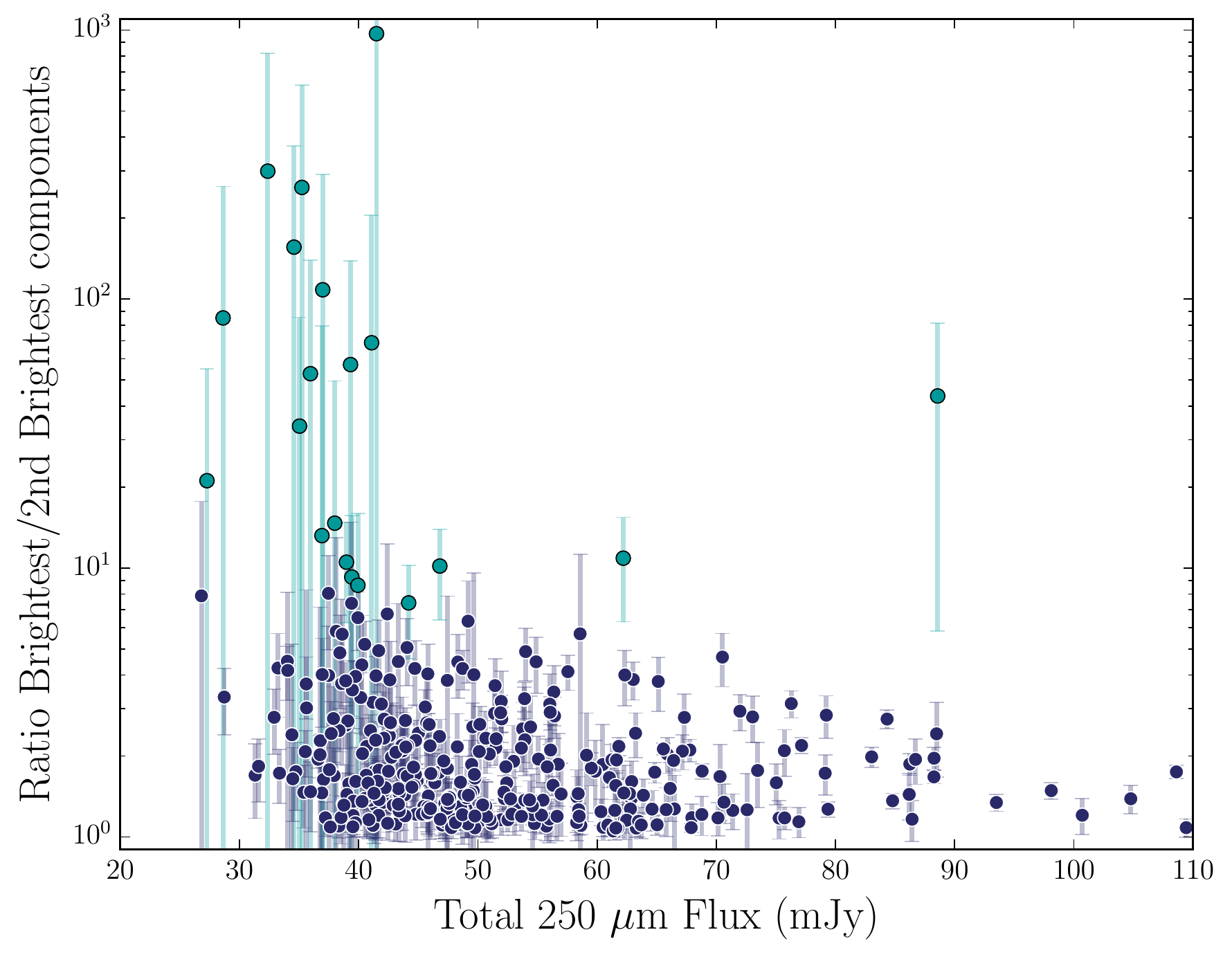}
   \caption{The ratio of the brightest component to the second brightest component in log, as a function of the total 250-\mm~source flux.  Values close to 1 indicate that the brightest component is nearly matched by the second brightest component.  Conversely, values much higher than 1 indicate that the brightest component is significantly brighter than the second brightest component. Dark blue points indicate that the second brightest component contains more than 10 per cent of the total 250-\mm~source flux.  Cyan points indicate that the second brightest component contains $<10$ per cent of the total 250-\mm~source flux. Vertical error bars on both sets of points are the standard deviation on the distribution of all possible ratios for the full 2000 samplings of brightest and second brightest components. A Spearman rank correlation test on the full sample of points plotted gives a p-value of $4.54\times10^{-9}$, which excludes the null hypothesis of no correlation at $>5\sigma$. A Spearman rank correlation test on the subsample with second brightest components containing $>10$ per cent of the total flux gives a p-value of $1.34\times10^{-5}$, which excludes the null hypothesis of no correlation at $>4\sigma$.}
   \label{fig:unbin_1to2}
\end{figure}

There is an additional consideration at this stage, which is that some fraction of our sample will have the majority of their flux assigned only to a single source, with extremely low fluxes found in the second brightest component.  
In order to consider more physically meaningful second components, we set an "insignificant flux'' threshold to 10 per cent of the total flux, which gives sources of  $>3$ mJy.
In Figure \ref{fig:unbin_1to2} we plot the full sample, but colour code according to the fraction of the total 250-\mm~source flux contained within the second brightest component. The dark blue points are those where the second brightest component has more than ten percent of the total source flux.  The cyan points indicate the 250-\mm~sources where the second brightest component contains less than 10 per cent of the total flux.  Our vertical error bars are the standard deviation on the full distribution of ratios, calculated by iterating through the full distribution of brightest and second brightest component fluxes.  These error bars therefore represent our full measurement uncertainty on each point.  We note that those data points plotted in dark blue typically have significantly smaller error bars, indicating that all iterations of the code have arrived at very similar flux values for the brightest and second brightest components, and that their probability distributions (as shown in Figure \ref{fig:triangle}) are narrow.  We emphasise that the points plotted are the medians of each distribution, for each 250-\mm~source.

As implied by Figures \ref{fig:bin_maxfraction} \& \ref{fig:bin_maxflux}, the majority of sources with very high brightest/second brightest component ratios are found at low 250-\mm~flux.
As the total flux of the 250-\mm~source increases, the maximal ratio between brightest and second brightest components decreases and approaches unity (i.e., the brightest and second brightest components are approximately equal in brightness).  At the low total source flux end, some of the brightest components are 1000 times brighter than their next brightest component; however, these extremes are present only for those 250-\mm~sources where the second component has insignificant flux.  This extreme difference between first and second brightest component also seems to dominate only where the total flux is very low (i.e., $< 40 $ mJy); above this level the brightest component tends to be no more than ten times brighter than the next brightest component. In this flux range ($>40$ mJy), components span the range from equally bright as their secondary component to several times brighter.  
A Spearman Rank correlation test (which is not sensitive to the specific shape of the correlation) on the total sample finds a correlation coefficient of $-0.3$, with a p-value of $4.54\times10^{-9}$, which excludes the null hypothesis at $>5\sigma$.  
The same test run on only the subsample with strong second brightest components (those points in dark blue) gives a p-value of $1.34\times10^{-5}$, which excludes the null hypothesis of no correlation at $>4\sigma$.

\begin{figure}
   \centering
   \includegraphics[width=250px]{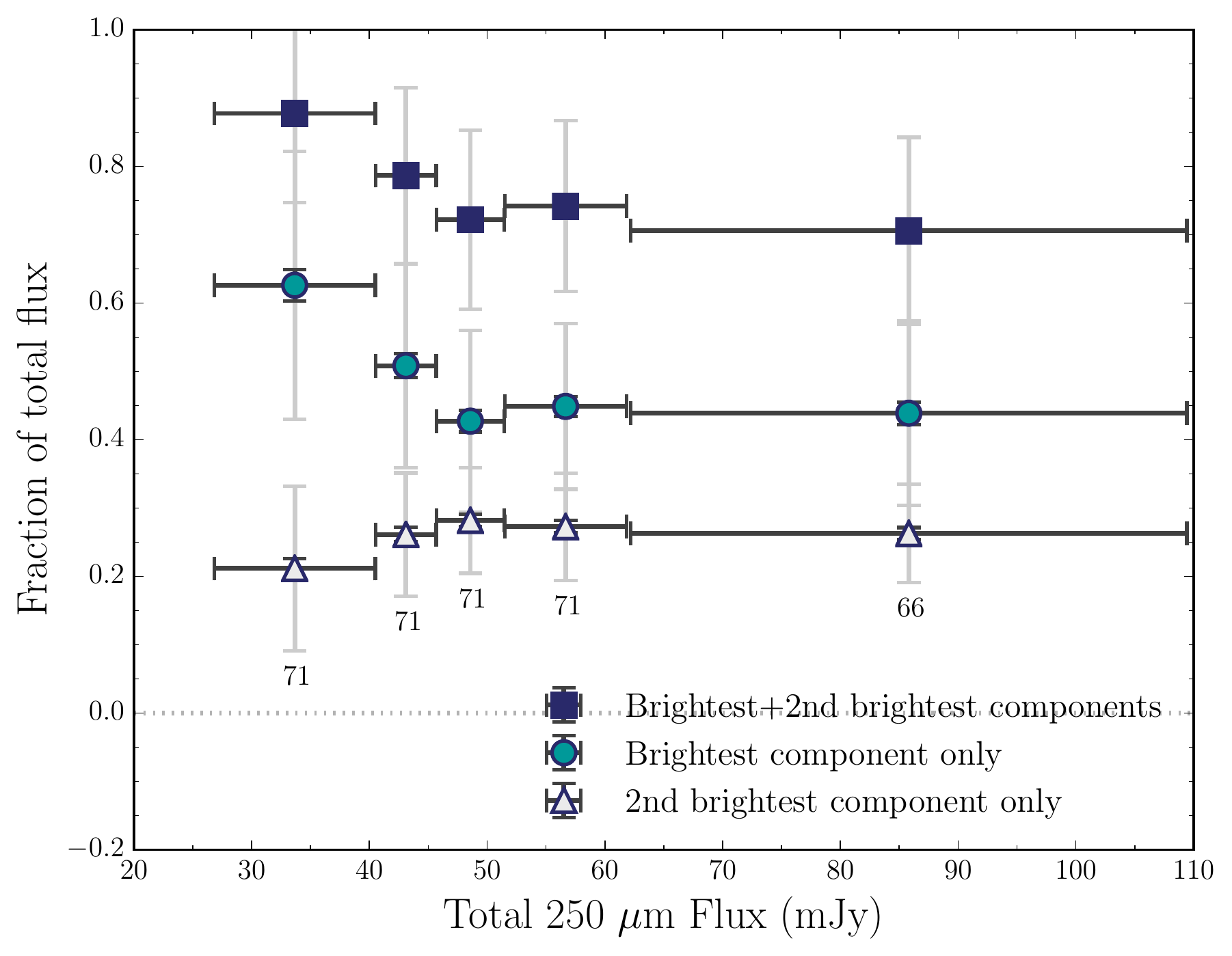}
   \caption{This Figure is constructed identically to Figure \ref{fig:bin_maxfraction}. We plot the brightest component's fraction of the total flux, as a function of the total flux, in cyan circles.  In grey triangles, we plot the second brightest component's contribution to the total flux. In dark blue squares, we plot the fraction of the total flux contained in the brightest and the second brightest components combined.  
    At the low total 250-\mm~flux end, the combination of the top two components accounts for approximately 90 per cent of the flux in the 250-\mm~source.  At the high total flux end, where we expect the brightest and second brightest components to be of nearly equal brightness, the top two sources account for approximately 70 per cent of the total flux.  At intermediate total fluxes, the sum of the brightest two components is reasonably constant at 70 per cent of the total flux.}
   \label{fig:bin_1and2}
\end{figure}

We plot the fraction of the total 250-\mm~source flux accounted for by the brightest two companions explicitly in Figure \ref{fig:bin_1and2}. As previously plotted in Figure \ref{fig:bin_maxflux}, the fraction of the total flux due to the brightest component is shown in cyan circles; the second brightest component is plotted on its own in grey triangles.  The sum of the brightest two components is plotted in dark blue squares.  The top two brightest components contribute approximately 80 per cent of the total 250-\mm~source flux at the high 250-\mm~total flux end, with only 20 per cent of the total flux remaining to be produced by additional companions. 
The typical contribution of the brightest two components is remarkably consistent between 70--80 per cent. 
The exception to this is the lowest total flux bin, which is consistent with the brightest two components containing 100 per cent of the flux.  The second brightest component is also remarkably consistent, producing 20--25 per cent of the total 250-\mm~flux across the full 250-\mm~flux range.

We examine the fraction of all 250-\mm~sources where the second brightest component contains a very small fraction of the total flux.  If the second component contains very little flux, we can consider this source to be a single component.  
If we continue to use our limit of the second brightest component containing at least 10 per cent of the total 250-\mm~flux, we find that only 5 per cent of all 250-\mm~sources are excluded.  We therefore conservatively conclude that our sample has a multiplicity fraction of 95 per cent, although the exact value will depend on the fractional threshold imposed on the secondary component.

\section{Discussion}	
\label{sec:discussion}

It is difficult to compare the results presented here directly to other studies, since the majority of studies in the FIR are selected (and observed) at different wavelengths than the current work.  However, our results may be more directly compared in the context of the conclusions drawn from other studies, keeping in mind the variable sample size between studies. In general, our results are consistent with previous studies of multiplicity in FIR sources.  

Problems with correctly identifying a shorter wavelength counterpart of a source detected in the FIR or sub-mm regime have been longstanding issues in this field.
Both \citet{Hughes1998} and \citet{Downes1999} struggled with the optical counterpart selection of their sub-mm observations, finding multiple possible options within a reasonable radius.  \citet{Hughes1998} had a set of five sub-mm detected sources, with optical counterparts either invisible, or with up to five potential counterparts within the beam.  \citet{Downes1999} conducted deep follow-up imaging of the brightest of the \citet{Hughes1998} sources and proposed that the sub-mm source was associated with four components, but noted the presence of nine other sources within 6 arcsec of the sub-mm source.  This difficulty then pushed the need for high resolution observations in the FIR or nearby wavelengths; observing at the same wavelength would help determine the correct optical counterpart.

Early attempts to do so were somewhat hampered by low number statistics.  One of the earlier high resolution observations of sub-mm source multiplicity was undertaken at the Sub-Millimeter Array (SMA) by \citet{Wang2011c}, who observed two sub-mm sources and found that they each source divided into two or three components.  The conclusion drawn was that it was highly likely sub-mm sources were being misidentified by being assigned to the nearest optical counterpart, and that all sub-mm sources would need to be re-observed with high resolution capabilities to ensure that the correct counterpart was identified at other wavelengths.

Subsequent studies have been highly varied in sample size, selection wavelength, and in how the multiplicity fraction was determined.  Early studies of multiplicity were necessarily of only a few objects at a time \citep[e.g.,][]{Wang2011c}, but the majority of recent studies have been of samples of order 20--30 objects, much smaller than the sample of 360 presented in the current work. For instance, \citet{Chen2013b} had a sample of 20 sources observed by SCUBA-2 at 450-\mm~or 850-\mm\footnote{SCUBA-2 has a FWHM beam size of 14.5 arcsec at 850 \mm.}; they determine that 2--3 (10--15 per cent) may be multiple components, given multiple nearby IRAC or 20 cm radio counterparts.  This estimate was made by investigating offsets between the centroid of the FIR flux and the radio counterpart.  However, one might expect this method to underestimate the true multiplicity fraction, as any multi-component structure which had a strong component associated with the radio component, and a second component not associated with the radio flux, would not be counted.  \citet{Hezaveh2013} have a similar sample size to \citet{Chen2013b}, finding 4 galaxies (originally detected at 1.4 mm at the South Pole Telescope\footnote{The South Pole Telescope has a FWHM beam size of 1 arcmin at 1.4 mm.}), which divide into directly imaged multiple components (in this case, strong gravitational lenses) out of a sample of 20, when observed with ALMA at 860 \mm, implying a multiplicity fraction of at least 20 per cent.

The majority of recent studies have relied on direct detection of bright components with interferometric observations.  However, this method also has its limitations.  \citet{Smolcic2012} calculated that at least 15 per cent of their 28 objects detected with 870-\mm~observations with LABOCA on the APEX 12-m single-dish telescope\footnote{LABOCA has a FWHM beam size of 19 arcsec at 870 \mm.}) were blends of multiple components, as the multiple components were directly imaged by the IRAM Plateau de Bure facility.  A number of the objects from the single-dish data were completely undetected with the higher resolution data.  \citet{Smolcic2012} concludes that it is very plausible that these non-detections are the result of the single dish flux being subdivided, so that the individual components are below the flux limit of their survey, which has a $3\sigma$ limit of $\sim1.4$ mJy. We note that this $3\sigma$ limit is approximately half as bright as the conservative threshold we impose in Section \ref{sec:2ndbrightest} to denote very low flux levels. By counting the non-detections as having multiple components, the multiplicity fraction of \citet{Smolcic2012} increases to 40 per cent.  This flux threshold limitation is in place for all such direct imaging campaigns to determine the multiplicity fraction.
The majority of such campaigns have similar number statistics to the \citet{Smolcic2012} work; \citet{Bussmann2015} finds that of their 29 FIR-selected sources, 20 (69 per cent) are multi-component in nature, down to a 5$\sigma$ limit of 1 mJy.  
Similarly, \citet{Simpson2015} investigate a sample of 30 galaxies detected by the SCUBA-2 Cosmology Legacy Survey at 850 \mm with ALMA, and find that 60 per cent of their sample is resolved into more than one component above a 4$\sigma$ limit of $\sim1$ mJy, with 4 sources resolved into 3 or 4 components. 
\citet{Chen2016} similarly finds that bright 850-\mm~sources ($>4$ mJy) are split into multiple components at least 40 per cent of the time.
While these smaller studies have multiplicity fractions that are significantly higher, they are still prone to the loss of low-flux counterparts, which may partially explain the difference between these results of 60--70 per cent and our result of a multiplicity fraction of $>95$ per cent.

With the advent of ALMA, larger surveys such as the ALMA LABOCA ECDFS Submillimeter Survey (ALESS) have been able to study samples of over a hundred sources \citep{Hodge2013, Karim2013} in high resolution, which brings much larger number statistics than have previously been available. \citet{Hodge2013} find a multiplicity fraction of 35 per cent in a sample of 126 galaxies selected with LABOCA. 
This 35 per cent reflects the fraction which are directly detected above a $3\sigma$ threshold of $\sim$1.5 mJy; considering the non-detections, this value increases to 50 per cent, similar to the results of \citet{Smolcic2012}.  This is, however, a significantly lower fraction than found by \citet{Simpson2015} and \citet{Bussmann2015}.  \citet{Simpson2015} point out that this discrepancy could well be due to the lower resolution of the ALESS sample relative to the \citet{Simpson2015} work, and that typically the single-dish fluxes of ALESS sources are lower than those of the sample probed in \citet{Simpson2015}.  Both of these aspects would make it more difficult to find lower-flux companions near a bright source, and explain the lower multiplicity fraction relative to that which is reported in the current work.  

An additional complication in comparing different studies is the increasing number of studies which suggest that the multiplicity fraction is dependent on the FIR or sub-mm source flux itself.  A number of studies have now shown evidence that the higher the total source flux, the more likely the source is to be divided into multiple components \citep{Hodge2013, Karim2013, Swinbank2014, Simpson2015}.  As \citet{Simpson2015} looked at higher flux objects than the ALESS team, one might expect them to find a higher multiplicity fraction.  In particular, \citet{Karim2013} find that above a given flux limit, \textit{all} sources are multiple components.   \citet{Cowley2015} present theoretical work which also suggests that bright sources ($>5$ mJy) have the majority of their flux (90 per cent) contained within 3--6 components, but that this multiplicity declines as source flux declines.  Similarly, another theoretical work \citep{Hayward2013} finds that the vast majority of sub-mm sources should be blends of multiple components; above a synthesized 850-\mm~flux of 3 mJy, the sub-mm sources are dominated by at least 2 physical components. The findings of \citet{Hayward2013}  are consistent with our results; the 250-\mm~sources at the faintest fluxes are most likely to be dominated by a single component, with the highest fluxes more likely to be divided into roughly equal flux components. 

Of the existing studies, only a few have begun to examine the properties of the resolved components.  \citet{Simpson2015} is one such, finding that the brightest component 
accounts for approximately 80 per cent of the total sub-mm flux, and that where there are 2 or more components, the secondary source tends to add approximately 25 per cent to the total flux.
The brightest component in the simulations presented in \citet{Cowley2015} is 70 per cent of the total flux on average.  This is significantly higher than we see here; typically our brightest component is 45 per cent of the total source flux, except at the lowest flux end, where the brightest component contains 60 per cent of the total flux.  Consistent with \citet{Simpson2015} observational study, we find that the secondary component has a fairly consistent contribution to the total flux, of about 20--25 per cent across all 250-\mm~source fluxes.  However, we typically reach fractions of 70--80 per cent only when considering the brightest two components jointly.  The \citet{Simpson2015} sources are selected at 850 \mm, so it is unclear how bright these sources would appear at 250 \mm. If the \citet{Simpson2015} study had investigated sources similar to those found within our lowest 250-\mm~flux bins, a brightest component fraction of 80 per cent would be close to what we observe. 

With the measurements of individual 250-\mm~fluxes associated with each 3.6-\mm~source, we can naturally extend this work to an analysis of whether or not the bright components are truly physically associated or not. Existing theoretical works suggest that the bright components are unlikely to be physically associated \citep{Hayward2013, Cowley2015, Munoz2015} except in a small fraction of cases. However, as the goal of this first paper is to determine how many of the 250-\mm~sources might be expected to divide into at least two components, a detailed observational analysis is beyond the scope of the current paper, and we defer it to to a forthcoming work.
Furthermore, our sample is typically divided into more counterparts than are traced by the 24-\mm~component (see also \citealt{Hodge2013}). Number counts derived only from 24-\mm~positions \citep[e.g.,][]{Bethermin2012} will likely be affected by the findings presented in this work.  Given that multiplicity changes as a function of total 250-\mm~flux density, it is not straightforward to estimate the exact influence; we also leave this to a future work.

\section{Conclusions}
\label{sec:conclusions} 
We have undertaken an analysis of a sample of 360 different 250-\mm~sources above $30$ mJy, found within the COSMOS field, which are selected to have at least one 24-\mm~counterpart and at least one 3.6-\mm~counterpart within the \herschel~beam.  We use the {\textsc{xid+}} code, which statistically determines all possible allocations of 250-\mm~flux based on the positions of the potential 24- and 3.6-\mm~components. Our conclusions can be summarised as follows.
\begin{itemize}
\item We find that while there is no statistically significant difference in the raw number of potential counterparts at 3.6- and 24-\mm~as a function of 250-\mm~source flux, the brightest and faintest 250-\mm~sources have statistically distinct distributions of flux to their respective components.

\item At low 250-\mm~source fluxes ($30 - 45$ mJy), the brightest components are assigned a larger fraction of the total 250-\mm~source flux. Above a 250-\mm~source flux of 45 mJy, the brightest components only contain $\sim 45$ per cent of the total 250-\mm~source flux.

\item The brightest components of the most luminous 250-\mm~sources are typically assigned more flux than the brightest components of faint 250-\mm~sources.

\item The most luminous 250-\mm~sources typically have brightest and second brightest components which are similar in flux; fainter 250-\mm~sources more often contain components which are very unequal in flux.

\item If both the brightest and second brightest components are considered, at the low 250-\mm~flux end ($<45$ mJy), 90 per cent of the 250-\mm~source flux is accounted for, indicating that only one or two sources are required to assemble the observed 250-\mm~flux.  At larger 250-\mm~fluxes, the two brightest components contain only 70 per cent of the total flux, indicating that further components are contributing significant flux.

\item Only 20/360 (5.6 per cent) have a secondary component that contains less than 10 per cent of the the 250-\mm~flux, indicating that $\sim 95$ per cent of our sample can be considered to be divided into multiple strong components.
\end{itemize}

\section*{Acknowledgments}
The authors thank the anonymous referee for a careful and very constructive report.
JMS thanks Steven Wilkins, Andreas Efstathiou, and Martin Harwit for useful comments on a draft of this manuscript, which helped to improve the clarity of this work.

JMS \& SJO acknowledge support from the Science and Technology Facilities Council (grant numbers ST/L000652/1).

JLW is supported by a European Union COFUND/Durham Junior Research Fellowship under EU grant agreement number 267209, with additional support from STFC (ST/L00075X/1).

The research leading to these results has received funding from the Cooperation Programme (Space) of the European Union's Seventh Framework Programme FP7/2007--2013/ under REA grant agreement number [607254].

The {\it Herschel} spacecraft was designed, built, tested, and launched under a contract to ESA managed by the Herschel/Planck Project team by an industrial consortium under the overall responsibility of the prime contractor Thales Alenia Space (Cannes), and including Astrium (Friedrichshafen) responsible for the payload module and for system testing at spacecraft level, Thales Alenia Space (Turin) responsible for the service module, and Astrium (Toulouse) responsible for the telescope, with in excess of a hundred subcontractors.

SPIRE has been developed by a consortium of institutes led by Cardiff Univ. (UK) and including: Univ. Lethbridge (Canada); NAOC (China); CEA, LAM (France); IFSI, Univ. Padua (Italy); IAC (Spain); Stockholm Observatory (Sweden); Imperial College London, RAL, UCL-MSSL, UKATC, Univ. Sussex (UK); and Caltech, JPL, NHSC, Univ. Colorado (USA). This development has been supported by national funding agencies: CSA (Canada); NAOC (China); CEA, CNES, CNRS (France); ASI (Italy); MCINN (Spain); SNSB (Sweden); STFC, UKSA (UK); and NASA (USA).

    This research has made use of data from HerMES project (http://hermes.sussex.ac.uk/). HerMES is a Herschel Key Programme utilising Guaranteed Time from the SPIRE instrument team, ESAC scientists and a mission scientist.

    The HerMES data was accessed through the Herschel Database in Marseille (HeDaM - http://hedam.lam.fr) operated by CeSAM and hosted by the Laboratoire d'Astrophysique de Marseille.

This work is based in part on observations made with the \textit{Spitzer Space Telescope}, which is operated by the Jet Propulsion Laboratory, California Institute of Technology under a contract with NASA.

\bibliographystyle{mnras}
\bibliography{masterfile_bibdesk}

\bsp	
\label{lastpage}
\end{document}